\documentstyle[12pt]{article}
\def\beq{\begin{equation}}
\def\eeq{\end{equation}}
\def\beqa{\begin{eqnarray}}
\def\eeqa{\end{eqnarray}}

\begin{document}
\begin{titlepage}
\begin{flushright}   UMDGR--96--036; 
		     arch-ive/9603166\\
\end{flushright}
\begin{center}
   \vskip 3em 
   {\LARGE Quantization of Scalar Field in the Presence of Imaginary 
	   Frequency Modes }
   \vskip 1.5em
   {\large Gungwon Kang\footnote{kang@rri.ernet.in} 
   \\[.5em]}
{\em Raman Research Institute, Bangalore 560 080, 
India\footnote{Present address}}  \\
and   \\
{\em Department of Physics, University of Maryland, College Park, 
MD 20742, USA}\\[.7em]
\end{center}
\vskip 1em
\begin{abstract}

Complex frequency modes occur for a scalar field near a rapidly rotating 
star {\it with ergoregion but no event horizon}. Such complex frequency modes 
must be included in the quantization of the field. As a model for this system, 
we have investigated a real scalar field with mass $\mu $ in a one-dimensional
square-well potential. If the depth of the potential is greater than $\mu^2$,
then there exist imaginary frequency modes. It is possible to quantize this
simple system, but the mode operators for imaginary frequencies satisfy 
unusual commutation relations and do not admit a Fock-like representation or 
a ground state. Similar properties have been discussed already by Fulling for 
a complex scalar field interacting with an external electrostatic potential. 

We are interested in the field dynamics in the physical case where the
initial state of the quantum field is specified before the complex 
frequency modes develop. As a model for this, we investigated a free 
scalar field whose ``mass" is normal in the past and becomes ``tachyonic"  
in the future. A particle detector in the far future placed in the 
in-vacuum state shows non-vanishing excitations related to the imaginary 
frequency modes as well. Implications of these results for the question 
of vacuum stability near rapidly rotating stars and possible 
applications to other fields in physics are discussed briefly.

\end{abstract}
\end{titlepage}
\newpage
\section{Introduction }
  
It has for a long time been known that the ergosphere of a rotating black 
hole leads to superradiance\cite{super}. The quantum counterpart of this
classical phenomenon of wave amplification, the so-called Starobinskii-Unruh
process\cite{Qsuper,Unford}, has been studied, predating Hawking's discovery
of the black hole evaporation\cite{Hawking}. It, however, is still 
unresolved whether the Starobinskii-Unruh effect is primarily due to the 
existence of the event horizon or the ergoregion.

To resolve this question, Matacz, Davies and Ottewill\cite{MDO} have 
investigated the quantum vacuum stability of a real massless scalar field 
near rapidly rotating stars that have {\it an ergoregion but no event 
horizon}. It turned out that the Starobinskii-Unruh effect is {\it absent}
in this setting if one assumes only real frequency modes occur. This result
may indicate that presence of the ergosphere is not sufficient for particle
production. However, as mentioned by the authors in Ref.\cite{MDO}, the
inclusion of complex frequency modes could change the conclusion seriously. 
In fact, Ashtekar and Magnon\cite{AM} have given a general argument based
on complex structure approach indicating particle production
near a star with ergoregion.  
In the case of rotating black holes, Whiting\cite{Whiting} has proved, for
massless fields, that complex frequency modes do not occur.  
On the other hand, in the case of rapidly rotating stars with ergoregions, 
negative energy could be accumulated within the ergoregion with giving 
radiation of positive energy to infinity. This process indeed generates
complex frequency modes\cite{vilen}; see the appendix for details. 
These modes are {\it exponentially}
amplified, reminiscent of a laser, form a discrete set, and may give rise to 
a novel form of vacuum instability after being quantized. The issue of the
ergoregion instability in stars has also been studied by many authors
from various points of view classically\cite{ergoinst}, although its quantum
counterpart has not as yet been understood well. 
Therefore, in order to conclude whether or not the vacuum instability occurs
near stars with ergoregions, one has to include these complex frequency modes 
as well and needs to understand their physical role in the problem. 
The first step will be to learn how to quantize complex frequency modes in
general. In addition, these complex frequency modes are also relevant in 
other physical systems such as leaky optical cavities\cite{LSL}, 
electromagnetic waves in plasma\cite{plasma}, $\alpha$-decay and the 
associated quantum mechanical tunnelling problems\cite{NW}, compound nucleus
theory\cite{WR}, and wave propagation in gravitational systems\cite{SO}. 

The quantization of a charged field in some electrostatic potentials 
including complex frequency modes has been briefly mentioned first by
Schiff, Snyder and Weinberg after they discovered the occurrence of those 
modes -- the so-called Schiff-Snyder-Weinberg effect --   \\
when the potential becomes strong\cite{ssw}. 
Schroer and Swieca\cite{schwieca} have constructed a {\it formal } 
quantization of a charged Klein-Gordon field with strong stationary external
interactions. This quantization of field including such complex frequency 
modes shows that there is no Fock-like representation, e.g., no normalizable
energy eigenstates, a breakdown of the vacuum postulate, and a breakdown of 
the particle interpretation of the quantum field theory. 
One specific and simple application of the above formulation was demonstrated 
by Schroer\cite{tachyon} for a free scalar field with a ``tachyonic" mass, 
i.e., $\mu^2 < 0$. Here the author shows how the inclusion of imaginary
frequency modes gives a consistent quantum field theory which is 
relativistically causal. 
Fulling\cite{Fulling} made clear the precise relationship of the occurrence 
of complex frequency modes to the Klein paradox\cite{Kpara}. It has also 
been shown how particle creation near a rotating {\it black hole} can be 
understood in terms of the Klein paradox\cite{Fulling,Manogue}, 
assuming complex frequency modes do not occur\cite{note}. Similarly, 
the quantum instability near a rotating {\it star}, if it occurs, may be
understood in terms of Schiff-Snyder-Weinberg effect. 

In the present paper, we describe the general features of quantization 
in the presence of complex frequency modes by considering a real scalar
field interacting with external scalar potential. The direct application
to the system of a star with ergoregion will appear elsewhere. 
In Sec.\ref{QPIFM}, the quantization is carried out in the presence of 
imaginary frequency modes. Many peculiar changes, due to imaginary frequency 
modes, in the conventional formulation of second-quantization are also shown
explicitly. In Sec.\ref{SM}, two simple models for potentials revealing 
imaginary frequency modes are given. Possible implications for the
issue\cite{MDO} of vacuum stability near rotating stars and applications 
to other fields in physics are discussed in Sec.\ref{disc}. 
In the appendix, the occurrence of complex frequency modes is reviewed for 
the case of rapidly rotating stars.    

\section{Quantization in the Presence of Imaginary     
\newline Frequency Modes}
\label{QPIFM}

As seen in the appendix, it is essential to include complex frequency modes 
as well as real ones in the quantization of fields near rapidly rotating 
stars. Instead of quantizing fields near stars with ergoregions directly, 
it will be much easier to consider a simpler system which contains the 
essential aspect of the problem. In this section, we describe a formulation 
of quantization for a real scalar field interacting with an external 
potential in Minkowski flat spacetime, assuming the potential is 
sufficiently negative so that imaginary frequency modes occur. 
Let us consider the following system in Minkowski flat spacetime:
\beq
{\cal L} = \frac{1}{2}(\partial_{\mu }\phi \partial^{\mu }\phi + 
\mu^2\phi^2 +V(x)\phi^2 )    
\label{Lag}
\eeq
where $\mu $ is the mass of the real scalar field $\phi (x)$ and $V(x)$ 
is an external scalar potential. The field equation is then
\beq
\frac{\partial^2}{\partial t^2} \phi (x)+ (-\nabla^2 +\mu^2 +V(x))
\phi (x) =0.     
\label{KGM}
\eeq
If the potential $V(x)$ is time-independent, solutions are separable 
in the form of  
\beq 
\phi_{(j)}(x) = \phi_{j}({\bf x}) e^{-i\omega_j t}     
\label{sep2}
\eeq
where $\phi_j({\bf x})$ satisfies
\beq
\omega_j^2\phi_j({\bf x}) = (-\nabla ^2 +\mu^2 +V({\bf x}))
\phi_j({\bf x}).
\label{sKG}
\eeq
This form of equation, Eq.~(\ref{sKG}), often appears for free fields 
in stationary curved spacetime as a radial part. For instance, see  
Eq.~(\ref{rKG}) in the appendix. Thus, by studying the simple system 
above, we are able to understand the essential behavior of fields in
curved spacetime as well. Eq.~(\ref{KGM}) is also analogous to that of
electromagnetic waves in an optical cavity with a position dependent 
dielectric constant $n^2({\bf x})$\cite{lcavity}. Before going further, 
let us consider the non-relativistic limit of the above equation. The
energy-momentum relation for a single particle corresponding to 
Eq.~(\ref{KGM}) will be $E^2=p^2+\mu^2+V(x)$, or $(E-\mu )(E+\mu )=p^2+V(x)$. 
For a non-relativistic particle $p<<\mu $ in a weak potential 
$V<<\mu^2$, $E\simeq \mu $ and so 
$E-\mu \simeq p^2/{2\mu }+(2\mu )^{-1}V(x)$, 
which is a form of Schr\"{o}dinger equation with a potential $(2\mu )^{-1}
V(x)$. For a strong potential 
$|V|>>\mu^2 $, one does not recover the usual form above and expects that 
$E$ becomes pure imaginary when the potential is sufficiently 
negative\cite{footnote}. 
 
Now note that the frequency $\omega_j$ in Eq.~(\ref{sKG}) could be either 
real or {\it pure imaginary} depending on the value of the potential 
$V({\bf x})$. That is, multiplying Eq.~(\ref{sKG}) by $\phi^{\ast }_j$ 
and integrating it, we find 
\beq
\omega^2_j = \frac{\int_{\Sigma }d^3x\, [|\nabla \phi_j|^2+(\mu^2 +
             V({\bf x}))|\phi_j|^2] -\oint_{\partial \Sigma }\phi^{\ast }_j
	     \nabla \phi_j \cdot d{\bf S} }{\int_{\Sigma }d^3x |\phi_j|^2}.
\label{freq}
\eeq
Here we assume suitable boundary behavior of the field so that the  
integration on the spatial boundary in Eq.~(\ref{freq}) vanishes\cite{BC}. 
Now it can be easily seen in Eq.~(\ref{freq}) that, if the potential 
$V({\bf x})$ is sufficiently negative, then there could exist mode solutions 
such that $\omega^2_j$ would be negative and so $\omega_j$ would be pure
imaginary, i.e., $\omega^{\ast }_j =-\omega_j$. In the case of free fields,
e.g., $V(x)=0$, the frequency must be real provided the boundary terms in
Eq.~(\ref{freq}) vanish. Note that even plane waves of complex frequencies 
with dispersion relations $\omega^2_j=k^2_j+\mu^2$ satisfy the KG equation.
However, the field becomes singular at spatial infinity, which cannot be
accepted physically in the free case.       

Now let us define a Klein-Gordon inner product induced from the field 
equation Eq.~(\ref{KGM}) \cite{Fulling}. Multiplying Eq.~(\ref{sKG}) by
$\phi_k^{\ast }$ and subtracting the same form for $\phi^{\ast }_k$ 
multiplied by $\phi_j$, we obtain
\beq
(\omega_j - \omega^{\ast }_k)(\omega_j +\omega^{\ast }_k)\int_{\Sigma }
d^3x \phi^{\ast }_k\phi_j = -\oint_{\partial \Sigma }(\phi^{\ast }_k
\nabla \phi_j - \phi_j \nabla \phi^{\ast }_k)\cdot d{\bf S}.
\label{inner1}
\eeq
Assuming suitable boundary conditions in which the right hand side of 
Eq.~(\ref{inner1}) vanishes, we see
\beq
(\omega_j - \omega^{\ast }_k)<\! \phi_k~,~\phi_j\! > =0
\label{inner2}
\eeq
where $<\! \phi_k\, ,\, \phi_j\! >=(\omega_j +\omega^{\ast }_k)\int d^3x \,  
\phi^{\ast }_k\phi_j =i\int_{t=0}d^3x\, \phi^{\ast }_{(k)}\! \stackrel{
\leftrightarrow }{\partial_t}\! \phi_{(j)}$. 
Since $\int d^3x\, \phi^{\ast }_{(k)}\! \stackrel{\leftrightarrow }
{\partial_t}\! \phi_{(j)}$ is time-independent, we define a Klein-Gordon 
inner product as follows:
\beq
<\! \phi_1\, ,\, \phi_2\! >= i\int_td^3x\, \phi^{\ast }_1\! 
\stackrel{\leftrightarrow }{\partial_t}\! \phi_2           
\label{inner3}
\eeq
at any space-like Cauchy surface.   
This equation yields non-trivial orthogonality relations 
in general--the so-called   
``quasi-orthogonality"\cite{Fulling,ssw}. First, notice from 
Eq.~(\ref{sep2}) and Eq.~(\ref{sKG}) that we have four linearly 
independent mode solutions 
\beq
\phi_{(j)}=(\omega_j , \phi_j)\, , \qquad  \phi^{\ast }_{(j)}= 
(-\omega^{\ast }_j , \phi^{\ast }_j)\, , \qquad 
\phi_{(\bar j)}=(\omega^{\ast }_j , \phi^{\ast }_j)\, , \qquad  
\phi^{\ast }_{(\bar j)}=(-\omega_j , \phi_j).
\label{msols}
\eeq
for given any mode solution $(\omega_j, \phi_j)\equiv \phi_j({\bf x})
e^{-i\omega_jt}$. If $\omega_j \neq \omega_k^{\ast }$, we find from
Eq.~(\ref{inner2}) that $\phi_{(j)}$ and $\phi_{(k)}$ are orthogonal:
\beq 
<\! \phi_{(k)}\, ,\, \phi_{(j)}\! > =0  \qquad  {\rm if} \qquad  
\omega_j \neq \omega_k^{\ast }.
\label{Qortho1}
\eeq
For real $\omega_j$, we define
\beq
\epsilon_j = <\! \phi_{(j)}\, ,\, \phi_{(j)}\! > =2\omega_j\! \int 
d^3x\, |\phi_j|^2.      \nonumber 
\eeq
For $\epsilon_j \neq 0 $, therefore, we can normalize $\phi_j$ so that 
$\epsilon_j =\pm 1$ according to the signature of the frequency $\omega_j$. 
For imaginary $\omega_j$, however, we see from Eq.~(\ref{Qortho1}) that
$\phi_{(j)}$ is {\it null}: $<\! \phi_{(j)}\, ,\, \phi_{(j)}\! > =0$.
But, $\phi_{(j)}$ and $\phi^{\ast }_{(\bar j)}$ are NOT orthogonal, and 
\beq
\bar \epsilon_j =<\! \phi^{\ast }_{(\bar j)}\, ,\, \phi_{(j)}\! >=
2\omega_j \int d^3x\, |\phi_j|^2       \nonumber 
\eeq
can be set to $\pm i$ according to the signature of $\hbox{Im}~\omega_j$
by normalizing $\phi_j$. According to Eq.~(\ref{Qortho1}), the pairs 
$(\phi_{(j)},\phi_{(\bar j)})$ and $(\phi^{\ast }_{(j)},
\phi^{\ast }_{(\bar j)})$ in Eq.~(\ref{msols}) are not necessarily 
orthogonal for real $\omega_j$, respectively, since $\omega_j=
\omega^{\ast }_k$. By linearly combining them, however, we can always find 
a set having the same form in which they are orthogonal. 
Hence we finally have the following ``quasi-orthogonality" relations
\beqa
\begin{array}{clcr}
<\! \phi_{(j)}\, ,\, \phi_{(k)}\! >=\epsilon_j \delta_{(j)(k)} \qquad 
\qquad  & \hbox{for real} \qquad \omega_j;  \nonumber      \\
<\! \phi_{(j)}\, ,\, \phi_{(j)}\! > =0, \qquad  <\! \phi^{\ast }_
{(\bar j)}\, ,\, \phi_{(j)}\! > = \bar \epsilon_j \qquad  
& \hbox{for imaginary} \qquad \omega_j  
\end{array}
\label{Qortho2}
\eeqa
where we can normalize the field $\phi_j$ so that $\epsilon_j=\pm 1$
and $\bar \epsilon_j=\pm i$ unless they vanish.

The total mode energy that the field mode contains classically is 
proportional to its norm. Thus, any imaginary frequency mode has zero 
total mode energy. 
To see this relationship it is convenient to introduce
the two-component formalism\cite{Fulling,Unford} of the Klein-Gordon
equations. The canonical conjugate $\pi (x)$ of the field $\phi (x)$
and the two-component field $\Phi (x)$ are defined as follows:
\beq
\pi = \frac{\partial L}{\partial \dot \phi }=\dot \phi ,\qquad \qquad 
\Phi = \left(\matrix{\phi \cr \pi \cr}\right).    \nonumber 
\eeq
Then Eqs.~(\ref{KGM}) and (\ref{sKG}) are respectively equivalent to 
\beq
i\frac{\partial }{\partial t}\Phi (t,{\bf x})=W\Phi (t,{\bf x}), \qquad
\qquad W\Phi_j({\bf x}) =\omega_j\Phi_j({\bf x})    \nonumber 
\eeq
where 
\beq
W=\left(\matrix{0&i\cr
		-i(-\nabla^2+\mu^2+V)&0\cr}\right)  , \qquad 
\Phi_j({\bf x})=\left(\matrix{\phi_j({\bf x}) \cr 
			      -i\omega_j\phi_j({\bf x}) \cr}\right).
\nonumber  
\eeq
Now we find from Eq.~(\ref{inner3}) that 
\beq
<\! \phi_1\, ,\, \phi_2\! >=-\int d^3x\, \Phi^{\dagger }_1\sigma_2\Phi_2 
\equiv <\! \Phi_1 \, ,\, \Phi_2 \! >     
\label{inner4}     
\eeq
where $\sigma_2=\left(\matrix{0&-i\cr i&0\cr}\right)$. Then we see
\beq
<\! \Phi_{(j)}\, ,\,W \Phi_{(j)}\! > =\int_{\Sigma }d^3x\, [|\pi_{(j)}|^2
+|\nabla \phi_{(j)}|^2+(\mu^2+V)|\phi_{(j)}|^2] -\oint_{\partial \Sigma } \, 
\phi^{\ast }_{(j)}\nabla \phi_{(j)} \cdot d{\bf S} .    
\label{energy}
\eeq
Therefore, if the mode solution satisfies suitable boundary conditions
such that the surface term above vanishes, then the total energy $H$ 
of the classical field mode, which is proportional to the first term 
in the above equation, 
will be proportional to the norm of the field mode,i.e.,
$\sim <\! \phi_{(j)}\, ,\, \phi_{(j)}\! >$. It implies that the total energy 
of a real frequency mode is $\sim \epsilon_j\omega_j$ whereas that of an
imaginary frequency mode is {\it zero}. However, the energy density is not
necessarily zero everywhere for imaginary frequency modes. 
In fact, one may easily show from Eq.~(\ref{energy}) that the mode energy 
density is exponentially increasing or decreasing, depending on the 
signature of Im $\omega_j$, in time and is negative in a region where the 
potential is strongly negative. 
In the case of a rapidly rotating star, the mode energy density for complex
frequencies is negative in the ergoregion which is exactly canceled by 
the positive energy outside as shown in the appendix. This property will 
also be demonstrated explicitly for the case of a square-well potential
in Sec.\ref{SWP}. Note also that $W$ is Hermitian with respect to the inner
product defined in Eq.~(\ref{inner4}). That is, assuming suitable boundary
conditions for the solutions, $<\! W\Phi_1\, ,\, \Phi_2\! >=<\! \Phi_1\, ,
\, W\Phi_2\! >$(For $\Phi_1=\Phi_2$, this relationship is always satisfied,
independent of boundary conditions.). Thus, the norm 
$<\! \Phi_1(t)\, ,\, \Phi_2(t)\! >$ is
time-independent and so is $<\! \phi_1(t)\, ,\, \phi_2(t)\! >$ 
as mentioned above.

Now the general solution of Eq.~(\ref{KGM}) will be 
\beqa
\phi (t,{\bf x})&=&\sum_{\omega_j >0}(a_jv_{(j)}+a^{\dagger }_j v^
{\ast }_{(j)}+ a_{\bar j}v_{(\bar j)}+a^{\dagger }_{\bar j}v^{\ast }
_{(\bar j)})
+\sum_{\rm Im ~\omega_j >0} (d_ju_{(j)}+d^{\dagger }_j u^{\ast }_{(j)}+
d_{\bar j}u_{(\bar j)}+d^{\dagger }_{\bar j}u^{\ast }_{(\bar j)})        
\nonumber    \\
&=&\sum_{\omega_j >0}(a_j\phi_je^{-i\omega_jt}+a^{\dagger }_j\phi^{\ast } 
_je^{i\omega_jt} + a_{\bar j}\phi^{\ast }_je^{-i\omega_jt}+a^{\dagger }
_{\bar j}\phi_je^{i\omega_jt})    \nonumber   \\
& & + \sum_{\rm Im ~\omega_j >0} [(d_j \phi_j+d^{\dagger }_j\phi^{\ast }_j) 
e^{-i\omega_jt} +(d_{\bar j}\phi^{\ast }_j+d^{\dagger }_{\bar j}
\phi_j) e^{i\omega_jt}]. 
\label{decom}
\eeqa
Note that the time dependence of mode solutions is exponentially 
increasing or decreasing ($\sim e^{\pm |\omega_j|t}$) for imaginary modes. 
To construct the quantum theory of this real scalar field we now 
interpret $\phi (t,{\bf x})$ as an operator-valued distribution with
the following equal-time commutation relations
\beq
\lbrack \phi (t,{\bf x})\, ,\, \pi (t,{\bf y})\rbrack = i\delta 
({\bf x - y})\, , \qquad 
\lbrack \phi (t,{\bf x})\, ,\, \phi (t,{\bf y}) \rbrack  = \lbrack \pi 
(t,{\bf x})\, ,\, \pi (t,{\bf y})\rbrack  =0  \nonumber    
\eeq                                          
where the canonical conjugate $\pi (t,{\bf x})=\partial \phi /{\partial t}
=\dot \phi $ is defined as usual. By using the following relations, 
\beq
a_j= <\! v_{(j)}\, ,\, \phi (x)\! >, \qquad  
d_j =-i<\! u^{\ast }_{(\bar j)}\, ,\, \phi (x)\! >, \qquad  
d_{\bar j}=i<\! u^{\ast }_{(j)}\, ,\, \phi (x)\! >,  \nonumber 
\eeq
we obtain commutation relations among mode operators:
\beqa
\lbrack a_j\, ,\, a^{\dagger }_k \rbrack =\delta_{jk}~, \qquad 
\lbrack a_j\, ,\, a_k\rbrack =\lbrack a^{\dagger }_j\, ,\, a^{\dagger }
_k\rbrack  =0 \qquad     \qquad      ;    \nonumber   \\
\lbrack d_j\, ,\, d_{\bar j}\rbrack =-i ~, \qquad \lbrack d_j\, ,
\, d^{\dagger }_j\rbrack =
\lbrack d_j\, ,\, d^{\dagger }_{\bar j}\rbrack =\lbrack d_{\bar j}
\, ,\, d^{\dagger }_{\bar j}\rbrack = 
\lbrack d_j\, ,\, d_k\rbrack =\lbrack d^{\dagger }_j\, ,\, d^{\dagger }_k
\rbrack = \cdots =0.
\label{comm}
\eeqa
Note the unusual form of commutation relations among mode operators 
for imaginary frequencies which result from the quasi-orthogonality 
properties among mode solutions. By ``unusual" we mean that $d_j$ and 
$d^{\dagger }_j$ commute but $d_j$ and $d_{\bar j}$ do not. 
It will be shown below that this property is related to the inverted harmonic 
oscillator representation of mode operators for imaginary frequencies.   
We may redefine mode operators for imaginary frequencies in a 
different way that $d'_j=e^{i\pi /4}d_j$ and $d'_{\bar j}=
e^{i\pi /4}d_{\bar j}$. Then, $\lbrack d'_j ~,~d'_{\bar j}\rbrack =
\lbrack d'^{\dagger }_j ~,~d'^{\dagger }_{\bar j}\rbrack =1$, 
and the rescaled mode functions 
$u'_{(j)}=e^{-i\pi /4}u_{(j)}$ and $u'_{(\bar j)}=e^{-i\pi /4}u_{(\bar j)}$
in Eq.~(\ref{decom}) will have the following orthogonality; 
$<\! u'^{\ast }_{(\bar j)}~, ~u'_{(j)}\! >=-i\bar \epsilon_j \to \pm 1$.

To check if these mode operators have particle interpretation let us 
consider the Hamiltonian operator for the Lagrangian Eq.~(\ref{Lag})
\beq
H=\frac{1}{2}\int d^3x\, [\pi^2 +(\nabla \phi )^2 +(\mu^2+V)\phi^2].
\label{Ham1}
\eeq
In terms of mode operators, we find
\beqa
H&=&\frac{1}{2}\sum_{\omega_j >0} \omega_j(a^{\dagger }_ja_j+a_ja^{\dagger }
_j+a^{\dagger }_{\bar j}a_{\bar j}+a_{\bar j}a^{\dagger }_{\bar j})  
\nonumber  \\
& &+ \frac{1}{2}\sum_{\rm Im~\omega_j >0} i\omega_j(d_{\bar j}d_j+
d_jd_{\bar j}
+d^{\dagger }_jd^{\dagger }_{\bar j}+d^{\dagger }_{\bar j}d^{\dagger }_j).
\label{Ham2}
\eeqa
Note $H^{\dagger }=H$ as expected. By linearly transforming into Hermitian
operators, e.g., $a_j=(\omega_j^{1/2}Q_j+i\omega_j^{-1/2}P_j)/{\sqrt{2}}$ 
with $\lbrack Q_j ~,~P_j\rbrack =i$, one can easily see that the Hamiltonian 
for real frequency modes has a representation of a set of {\it attractive}
harmonic oscillators as usual. Thus, the energy spectrum is discrete and 
bounded below. By defining a vacuum state such that $a_j|\, 0\, >=0$ for all
$j$, one can construct a Fock space which possesses the conventional 
particle interpretation. For the mode operators of imaginary frequencies,
let us consider following linear transformations,     
\beqa
d_j\! &=&-\frac{1}{2}[i(|\omega_j|^{1/2}q_j +|\omega_j|^{-1/2}p_j) 
      +(|\omega_j|^{1/2}q_{\bar j}-|\omega_j|^{-1/2}p_{\bar j})],
      \nonumber   \\
d_{\bar j}\! &=&\frac{1}{2}[i(|\omega_j|^{1/2}q_j-|\omega_j|^{-1/2}p_j)
      +(|\omega_j|^{1/2}q_{\bar j}+|\omega_j|^{-1/2}p_{\bar j})].
\label{harm}
\eeqa
Here $q$ and $p$ are Hermitian operators satisfying $\lbrack q\, ,
\, p\rbrack =i$ again. Then, the Hamiltonian operator for imaginary 
frequency modes can be expressed by  
\beqa
\sum_{\rm Im~\omega_j >0} H_j&=&\sum_{\rm Im~\omega_j >0} 
\frac{1}{2}i\omega_j(d_{\bar j}d_j+
d_jd_{\bar j}+d^{\dagger }_j
d^{\dagger }_{\bar j}+d^{\dagger }_{\bar j}d^{\dagger }_j) \nonumber  \\
&=&\sum_{\rm Im~\omega_j >0} \frac{1}{2}(p_j^2 -|\omega_j|^2q_j^2 + 
p^2_{\bar j}-|\omega_j|^2
q^2_{\bar j}).
\label{Ham3}
\eeqa       
We find that $H_j$ is a system of two decoupled {\it repulsive} 
harmonic oscillators
with frequency $|\omega_j|$. The energy spectrum is continuous then and 
not bounded below. Thus, it has no ground state. One cannot define a 
reasonable vacuum state for $H_j$. In fact, the Hamiltonian $H_j$ has no
normalizable eigenstate. We shall show an explicit construction of energy
eigenfunctions below which is non-normalizable. Suppose it has an energy
eigenstate such that $H|\psi_E\! >=E|\psi_E\! >$. Notice from 
Eq.~(\ref{comm}) and Eq.~(\ref{Ham2}) that
\beq
\lbrack H,d_j\rbrack =-\omega_j d_j, \qquad \lbrack H,d^{\dagger }_j\rbrack 
=-\omega_jd^{\dagger }_j, \qquad \lbrack H,d_{\bar j}\rbrack =\omega_j
d_{\bar j}, \qquad \lbrack H,d^{\dagger }_{\bar j}\rbrack =\omega_j
d^{\dagger }_{\bar j}. 
\eeq
Thus, $d_j|\psi_E\! > $ and $d^{\dagger }_j|\psi_E\! >$ and $d_{\bar j}
|\psi_E\! >$ and $d^{\dagger }_{\bar j}|\psi_E\! >$ are eigenstates with
eigenvalues $(E\pm \omega_j)$, respectively, which are {\it not real} any 
more since $\omega_j$ is imaginary. This fact is not inconsistent with
the Hermiticity of the Hamiltonian operator because energy eigenstates are 
not normalizable ones. All these properties described above, therefore, 
indicate that imaginary frequency mode operators do not possess a Fock-like
space although there is a Hilbert space for them as will be shown below. 

Before explicitly constructing a Hilbert space for $H_j$, one finds that
it is convenient to define mode operators in a different way, which also
makes the connection of imaginary mode operators to the inverted harmonic 
oscillator representation transparent. Let us first consider an operator
$H=\frac{1}{2}(p^2+\omega^2q^2)$ where $p$ and $q$ are Hermitian operators
satisfying $\lbrack q,p\rbrack =i$, and $\omega $ is an arbitrary 
{\it complex} number. 
Then, $H=\frac{1}{2}\omega [(\omega^{1/2}q)^2+(\omega^{-1/2}p)^2
]=\frac{1}{2}\omega (ab+ba)$ where $a=(\omega^{1/2}q+i\omega^{-1/2}p)/
\sqrt{2}$ and $b=(\omega^{1/2}q-i\omega^{-1/2}p)/\sqrt{2}$ satisfying
$\lbrack a,b\rbrack =1$. Note that $\lbrack H,a\rbrack =-\omega a$ and 
$\lbrack H,b\rbrack =\omega b$. For a real $\omega $ corresponding to an
attractive harmonic oscillator, $a^{\dagger }=b$ and so $H=\frac{1}{2}
\omega (aa^{\dagger }+a^{\dagger }a)$ with $\lbrack a,a^{\dagger }\rbrack 
=1$. For a pure imaginary $\omega =i|\omega |$, which corresponds to a 
repulsive harmonic oscillator, $a=\sqrt{i}(|\omega |^{1/2}q+|\omega |^{-1/2}
p)/\sqrt{2}=ia^{\dagger }$ and $b=\sqrt{i}(|\omega |^{1/2}q-|\omega |
^{-1/2}p)/\sqrt{2}=ib^{\dagger }\neq a^{\dagger }$.  
Let $\phi_j(x)=\phi^R_j(x)+i\phi^I_j(x)$ where $\phi^R_j$ and $\phi^I_j$
are real functions. Then, the field decomposition in Eq.~(\ref{decom})
has the following form 
\beq
\phi (t,{\bf x})=\cdots +(b'_j\phi^R_j+c'_j\phi^I_j)e^{-i\omega_jt}+
(b'_{\bar j}\phi^R_j+c'_{\bar j}\phi^I_j)e^{i\omega_jt}+\cdots 
\eeq
where
\beq
b'_j=d_j+d^{\dagger }_j=b'^{\dagger }_j~,~ b'_{\bar j}=d_{\bar j}+
d^{\dagger }_{\bar j}=b'^{\dagger }_{\bar j}~,~ c'_j=i(d_j-d^{\dagger }_j)
=c'^{\dagger }_j ~,~ c'_{\bar j}=-i(d_{\bar j}-d^{\dagger }_{\bar j})
=c'^{\dagger }_{\bar j}.
\label{real}
\eeq
Then, $\lbrack b'_j\, ,\, b'_{\bar j}\rbrack =\lbrack c'_j\, ,\, c'_{\bar j}
\rbrack =-2i$, otherwise vanishes. By redefining $b'_j=\sqrt{-2i}b_j$ 
and $c'_j=\sqrt{-2i}c_j$ so that 
$\lbrack b_j,b_{\bar j}\rbrack =\lbrack c_j,c_{\bar j}\rbrack =1$, 
the Hamiltonian operator $H_j$ in Eq.~(\ref{Ham3}) becomes  
\beq
H_j =\frac{1}{2}\omega_j(b_jb_{\bar j}+b_{\bar j}b_{j}+
c_jc_{\bar j}+c_{\bar j}c_{j}).
\eeq    
Now we can easily see that $H_j$ is equivalent to a system of two
decoupled {\it repulsive} harmonic oscillators with frequency $|\omega_j|$. 
As a consequence, one also finds that the unusual feature of commutation
relations in Eq.~(\ref{comm}) for imaginary mode operators is indeed related 
to a property of the {\it inverted} harmonic oscillator system.   
 
By defining a new set of Hermitian operators $x=\sqrt{-i}b_j, p_x=
-i\partial_x=-\sqrt{-i}b_{\bar j}, y=\sqrt{-i}c_j$, 
and $p_y=-i\partial_y=-\sqrt{-i}c_{\bar j}$, one can
easily find energy eigenfunctions\cite{schwieca}   
\beq
\psi_{\varepsilon m}(r,\varphi )=(2\pi )^{-1}r^{-i\varepsilon -1}
e^{im\varphi }   \nonumber 
\eeq
where $\varepsilon $ is an arbitrary continuous real number and $m$ 
an integer. Since $H_j\psi_{\varepsilon m}=\frac{1}{2}\varepsilon 
|\omega_j|\psi_{\varepsilon m}$, the continuous energy eigenvalue is
$E_j=\frac{1}{2}\varepsilon |\omega_j|$. These eigenfunctions are orthogonal
\beq
\int \psi^{\ast }_{\varepsilon 'm'}\psi_{\varepsilon m}rdrd\varphi 
=\delta (\varepsilon -\varepsilon ')\delta_{mm'},    
\label{nonnor}
\eeq
and are complete
\beq
\int^{\infty }_{-\infty }d\varepsilon \psi^{\ast }_{\varepsilon m}
({\bf x})\psi_{\varepsilon m}({\bf y})=\delta ({\bf x}-{\bf y}). 
\nonumber
\eeq
Note from Eq.~(\ref{nonnor}) that energy eigenfunctions are not normalizable.  
However, one can construct normalizable wave packets from them\cite{tachyon}.
These square integrable wave packets form a Hilbert space ${\cal H}_j$.    
The Hilbert space for the quantum field $\phi $ is then 
\beq
{\cal H}={\cal H}_{\rm Re} \otimes \prod_j {\cal H}_j    \nonumber
\eeq
where ${\cal H}_{\rm Re}$ is the usual symmetrized Fock space generated 
by real frequency modes and $\prod_j{\cal H}_j$ by imaginary frequency 
modes which can be either a finite number of products or an infinite number 
of products depending on how many imaginary frequency modes occur 
in the system. 

\section{Specific Models}
\label{SM}

As a simple model for the system, we investigate a real scalar field with 
mass $\mu$ in a one-dimensional static square-well potential, and 
explicitly show that a finite number of imaginary frequency modes appear 
as the depth of the square-well becomes deeper than $\mu^2$. In addition, 
we will also investigate a free scalar field in a time varying potential 
since we are interested in the physical case where the initial state of 
the quantum field is specified before the complex frequency modes develop. 
For instance, it will be very interesting to study the vacuum stability of 
quantum fields in the distribution of collapsing matter which finally 
settles down to a rapidly rotating star {\it with ergoregion}. 

\subsection{Square-well potentials}
\label{SWP}

\indent
Now let us apply the formulation constructed above to a specific model
in two-dimensional spacetime in which the potential is 
\beq
\begin{array}{clcr}
V(x)=\cases{0   \qquad & \mbox{for $|x| > a$},  \cr 
	    -V_0 \qquad & \mbox{for $|x| <a$}. \cr}
\end{array}
\eeq
Then, the field equation Eq.~(\ref{sKG}) becomes
\beqa
\begin{array}{ll}     
d^2\phi_j/{dx^2} +(\omega^2_j-\mu^2)\phi_j =0 \qquad 
& \mbox{for $|x|>a$},     \\
d^2\phi_j/{dx^2} +(\omega^2_j-\mu^2+V_0)\phi_j =0 \qquad 
& \mbox{for $|x|<a$}.
\end{array}
\eeqa
The solutions will be stationary or exponential depending on the sign 
of the coefficients of the second terms. As boundary conditions, we   
assume that $\phi_j(x)$ is not singular at $x=\pm \infty$, and that 
$\phi_j$ and $d\phi_j/{dx}$ are continuous at $x=\pm a$\cite{foot2}. 

If the frequency $\omega_j$ is complex, the argument below Eq.~(\ref{freq}) 
shows that it should be {\it pure imaginary}. If $V_0\leq \mu^2$, which 
also includes the case of a step potential when $V_0<0$, then both inside 
and outside solutions are exponential for {\it imaginary} frequencies. 
Since this class of solutions cannot satisfy the continuity of the first 
derivative at $x=\pm a$, there is no imaginary frequency mode solution in 
this case and so the quantization will end up with the Fock space 
${\cal H}_{\rm Re}$ only. For $V_0>\mu^2$, however, there are three classes 
of mode solutions including {\it imaginary} frequency ones. 
Let $\omega_j^2-\mu^2+V_0 = k^2$. Hence, $\omega_j^2-\mu^2=k^2-V_0$.  
Since $\omega_j$ is possibly pure imaginary only, $k^2$ must be real. 
If $k^2$ is negative(and so is $k^2-V_0$ as well), there will be 
no solutions satisfying continuity 
conditions at $x=\pm a$ by the same reason above. Thus, it is sufficient 
to consider only real $k$ ranging from $-\infty $ to $\infty $.     

(i) If $k^2>V_0$, then $\omega_j^2 >\mu^2$ and so $\omega_j$ is real. There 
are two linearly independent solutions $^R\phi_k(x)$ and $^L\phi_k(x)$ for
any {\it continuous} $k$ in this range:
\beq
^R\phi_k(x) =\left\{\begin{array}{ll}
                      e^{i\sqrt{k^2-V_0}\, x} +R_Re^{-i\sqrt{k^2-V_0}\, x} 
\qquad   & \mbox{for $x\leq -a$,}  \\ 
                      A_R\sin kx +B_R\cos kx 
         & \mbox{for $-a\leq x \leq a$,}   \\  
                      T_Re^{i\sqrt{k^2-V_0}\, x} 
         & \mbox{for $x \geq a$,}  
                    \end{array}
             \right. 
\eeq
and
\beq
^L\phi_k(x) =\left\{\begin{array}{ll}
                     T_Le^{-i\sqrt{k^2-V_0}x}  &  
\mbox{for $x\leq -a$,}    \\ 
                     A_L\sin kx +B_L\cos kx    &
\mbox{for $-a\leq x\leq a$,}    \\
e^{-i\sqrt{k^2-V_0}x} +R_Le^{i\sqrt{k^2-V_0}x}  \qquad  &
\mbox{for $x\geq a$.}  
                    \end{array}
             \right.  
\eeq
The coefficients are determined by the continuity conditions at $x=\pm a$.
$^R\phi_k(x)$ can be regarded as a right-going wave with reflection at 
$x=-a$. Similarly, $^L\phi_k(x)$ is a left-going wave with reflection at
$x=a$.     

(ii) If $V_0-\mu^2 \leq k^2 \leq V_0$, then $0\leq \omega_j^2 \leq \mu^2$ and
so $\omega_j$ is still real, but the outside solution must be exponential 
since $\omega^2_j-\mu^2=k^2-V_0<0$. There are even or odd solutions for 
some {\it discrete} values of $k$ in this range:
\beq
{^e}\phi_{k_j}(x) =\left\{\begin{array}{ll} 
                            e^{\sqrt{V_0-k^2_j}x}     & 
\mbox{for $x\leq -a$,}     \\
                            C_e\cos k_jx  \qquad \qquad  &
\mbox{for $-a\leq x\leq a$,}   \\
                            e^{-\sqrt{V_0-k^2_j}x}    &
\mbox{for $a\leq x$,}
                           \end{array}
                    \right. 
\label{esol}
\eeq
and
\beq
{^o}\phi_{k_j}(x)=\left\{\begin{array}{ll}
                           e^{\sqrt{V_0-k^2_j}x}   &
\mbox{for $x\leq -a$,}      \\  
                           C_o\sin k_jx   \qquad \qquad   &
\mbox{for $-a\leq x\leq a$,}   \\
                           -e^{-\sqrt{V_0-k^2_j}x}   &
\mbox{for $a\leq x$.}  
                          \end{array}
                   \right.
\label{osol}
\eeq
Here $k_j$ is determined by continuity conditions at $x=\pm a$ again. 
For even modes, 
\beq
k_ja\, \tan k_ja =\sqrt{V_0a^2-(k_ja)^2}.
\label{emode}
\eeq
Thus, depending on $\sqrt{V_0}a,\, \mu a$, and $a$, there could be no 
$k$ value giving a solution or a finite number of $k$'s which give mode 
solutions. For odd modes, 
\beq
k_ja\, \cot k_ja  =-\sqrt{V_0a^2-(k_ja)^2}. 
\label{omode}
\eeq

(iii) If $0\leq k^2\leq V_0-\mu^2$, then $-(V_0-\mu^2)\leq \omega^2_j <0$
and so $\omega_j$ is {\it imaginary}. The form of mode solutions is the 
same as in case (ii), i.e., Eq.~(\ref{esol})-(\ref{osol}). The equations 
determining the $k$'s are also the same as in 
Eq.~(\ref{emode})-(\ref{omode}), but the range of $ka$ is different. 
Since it includes zero, there always exists 
at least one even mode solution. 

We have shown before that the total classical mode energy is zero for 
imaginary frequency modes. One can check this property for any imaginary 
frequency mode solutions. For example, the energies stored outside and 
inside the square-well for an even mode are 
\beq
H_{\rm OUT} \sim \sqrt{V_0-k_j^2}e^{-2\sqrt{V_0-k_j^2}a}e^{2|\omega_j|t}, 
\qquad H_{\rm IN} \sim - k_jC^2_e\cos k_ja \sin k_jae^{2|\omega_j|t}.   
\nonumber 
\eeq
Note that both they are incresing exponentially in time. However,
the total energy is 
\beq
H=H_{\rm OUT}+H_{\rm IN} \sim (\sqrt{V_0-k^2_j}-k_j\tan k_ja)
e^{2|\omega_j|t}    \nonumber
\eeq
which vanishes because of Eq.~(\ref{emode}). Note also that $H_{\rm IN}$
is always {\it negative} for imaginary modes as expected from the form 
of the energy density in Eq.~(\ref{Ham1}). 

The field operator in Eq.~(\ref{decom}) can be written as 
follows in this case:
\beqa
\phi (t,x)&=& \sum_{k^2>V_0 (k>0)}\, [({^R}a_k{^R}\phi_ke^{-i\omega t}
+{^R}a^{\dagger }_k{^R}\phi_k^{\ast } e^{i\omega t})+({^L}a_k{^L}
\phi_ke^{-i\omega t}+{^L}a_k^{\dagger }{^L}\phi_k^{\ast }e^{i\omega t})]   
\nonumber  \\
& & +\sum_{V_0-k^2\leq k^2_j\leq V_0}\, [({^e}a_{k_j}{^e}\phi_{k_j}
e^{-i\omega_jt}+{^e}a^{\dagger }_{k_j}{^e}\phi_{k_j}^{\ast }e^{i\omega_jt})
+({^o}a_{k_j}{^o}\phi_{k_j}e^{-i\omega_jt}+{^o}a^{\dagger }_{k_j} 
{^o}\phi_{k_j}^{\ast }e^{i\omega_jt})]        \nonumber    \\
& & +\sum_{0\leq k^2_j<V_0-\mu^2}\, [({^e}d_{k_j}{^e}\phi_{k_j}
+{^e}d^{\dagger }_{k_j}{^e}\phi_{k_j}^{\ast })e^{-i\omega_jt}
+({^o}d_{k_j}{^o}\phi_{k_j}+{^o}d^{\dagger }_{k_j}{^o}\phi_{k_j}^{\ast })
e^{i\omega_jt}]    
\eeqa
where $\omega =+\sqrt{k^2+\mu^2-V_0}$ for real frequencies and 
$\omega_j=+i\sqrt{V_0-(k^2_j+\mu^2)}$ for imaginary frequencies. Note that 
$^R\phi_k(x)e^{-i\omega t}$ and $^L\phi_k(x)e^{-i\omega t}$ become 
purely right-going and left-going waves,i.e., $e^{ikx-i\omega t}$ and
$e^{-ikx-i\omega t}$, in the limit of $V_0\to 0$, respectively.   
Now the quantization of this field simply follows the general scheme 
described in Sec.\ref{QPIFM}.   

\subsection{Time-varying mass}    
\label{TVM}

\indent   
Instead of studying an ``eternal" rotating star with ergoregion, it is 
more realistic to consider a dynamical rotating star system. That is, 
the spacetime was almost flat in the past with some matter distribution 
possessing non-zero total angular momentum. This matter starts to 
collapse and finally settles down to a rapidly rotating star having  
ergoregion. In this situation, we are mainly interested in the field 
dynamics in which the state of the quantum field in the past is specified
(for instance, the in-vacuum state) before the complex frequency modes
develop. As a model for this, we shall investigate below a time-varying 
potential $V(t)$ such that
\beq
V(t)=-\frac{V_0}{2}(\tanh \rho t+1)=\left\{\begin{array}{ll}
                                             0 \qquad \qquad   &
\mbox{as $t \to -\infty $,}     \\
                                             -V_0      &
\mbox{as $t \to \infty $.}
                                           \end{array}
                                    \right.
\eeq
Separating $\phi (t,x)=e^{ikx}\phi_k(t)$, we find that Eq.~(\ref{KGM})
yields
\beq
\frac{d^2\phi_k}{dt^2} +(k^2+\mu^2C(t))\phi_k =0
\label{tKG}
\eeq
where
\beqa
C(t)&=&1+\mu^{-2}V(t)=(1-V_0/{2\mu^2})-\frac{V_0}{2\mu^2}\tanh \rho t 
\nonumber   \\
&=& \left\{\begin{array}{ll}
             1                 &
\mbox{as $t \to -\infty $,}    \\
             -\mu^{-2}(V_0-\mu^2) \qquad \qquad   &
\mbox{as $t \to \infty $.}  
	   \end{array}
     \right. 
\nonumber   
\eeqa
For $V_0 >\mu^2$, the system is equivalent to a free scalar field starting
with normal mass $\mu^2$ and ending up with ``tachyonic" mass 
$-(V_0-\mu^2)<0$. Therefore, imaginary frequency modes occur as time 
evolves. For example, in the far future $t \sim \infty $, all modes 
satisfying $k^2+\mu^2<V_0$ have imaginary frequencies $\omega (k)=\pm 
i\sqrt{V_0-(k^2+\mu^2)}$. Schroer\cite{tachyon} has already studied the
quantization of a scalar field with tachyonic mass, showing one of 
remarkable results that the tachyons propagate {\it causally}. 

Eq.~(\ref{tKG}) can be solved exactly in terms of hypergeometric functions.  
The mode solutions which behave like plane waves with positive frequency 
$\omega_{\rm in}=\sqrt{k^2+\mu^2}$ in the remote past are
\beqa
\phi^{\rm in}_k(t) &\sim & e^{-i\omega_+t-i\omega_-/\rho \ln (2\cosh 
\rho t)}
{_2}F_1(1+i\omega_-/\rho ,i\omega_-/\rho ;1-i\omega_{\rm in}/\rho ;
(1+\tanh \rho t)/2)   \nonumber   \\
&\longrightarrow & e^{-i\omega_{in}t} \qquad  \qquad \hbox{as} 
\qquad t\to -\infty .
\nonumber 
\eeqa
On the other hand, the modes which behave like plane waves with 
frequency $\omega_{\rm out}=   \\
\sqrt{k^2+\mu^2-V_0}$ in the far future are
\beqa
\phi^{\rm out}_k(t)&\sim &e^{-i\omega_+t-i\omega_-/\rho \ln (2\cosh \rho t)}
{_2}F_1(1+i\omega_-/\rho ,i\omega_-/\rho ;1+i\omega_{\rm out}/\rho ;
(1-\tanh \rho t)/2)   \nonumber   \\
&\longrightarrow & e^{-i\omega_{\rm out}t} \qquad \qquad \hbox{as} \qquad
t\to \infty       \nonumber
\eeqa
where $\omega_{\pm }=(\omega_{\rm out}\pm \omega_{\rm in})/2$. Thus, the 
asymptotic behavior of $\phi^{\rm out}_k$ for $k^2+\mu^2 < V_0$ is that of 
imaginary frequency modes of $\omega_{\rm out}=i\sqrt{V_0-(k^2+\mu^2)}$. 
These two complete bases lead to the following two equivalent expansions 
of the field $\phi (t,x)$ 
\beq
\phi (t,x)=\sum_{k}\, (a_ku^{\rm in}_k +a^{\dagger }_k
u^{\rm in\ast }_k) =\sum_{k}\, (b_ku^{\rm out}_k+b^{\dagger }_k
u^{\rm out\ast }_k)   
\label{tdecom}
\eeq
where $u_k(t,x)=e^{ikx}\phi_k(t)$. Since the Klein-Gordon inner product
is time-independent for solutions satisfying suitable boundary conditions 
at $x=\pm \infty $, we obtain
\beq
<\! u^{\rm in}_{k'}\, ,\, u^{\rm in}_k\! >|_{t=t} = <\! u^{\rm in}_{k'}
\, ,\, u^{\rm in}_k\! >|_{t=-\infty } =\epsilon_k \delta_{k'k},  \nonumber
\eeq
and
\beqa
\begin{array}{ll}
  <\! u^{\rm out}_{k'}\, ,\, u^{\rm out}_k\! >|_{t=t} = <\! u^{\rm out}_
{k'}\, ,\, u^{\rm out}_k\! >|_{t=\infty } =\epsilon_k \delta_{k'k} 
\qquad    & \mbox{for $k^2+\mu^2>V_0$,}   \\  
  <\! u^{\rm out\ast }_{-k}\, ,\, u^{\rm out}_k\! >=-i, \qquad 
  <\! u^{\rm out}_{k}\, ,
  \, u^{\rm out}_k\! >=0     & \mbox{for $k^2+\mu^2<V_0$.} 
\end{array}
\nonumber 
\eeqa
Thus, the commutation relations among mode operators are
\beqa
\begin{array}{ll}
  \lbrack a_k\, ,\, a^{\dagger }_{k'}\rbrack =\delta_{kk'} ~, ~
  \lbrack a_k\, ,\, a_{k'}\rbrack =\lbrack a^{\dagger }_k\, ,\, 
  a^{\dagger }_{k'}\rbrack = 0 \qquad  &
\mbox{for all $k$,}   \\  
  \lbrack b_k\, ,\, b^{\dagger }_{k'}\rbrack =\delta_{kk'} ~, ~
  \lbrack b_k\, ,\, b_{k'}\rbrack =\lbrack b^{\dagger }_k\, ,\, 
  b^{\dagger }_{k'}\rbrack = 0    & 
\mbox{for $k^2+\mu^2>V_0$,}     \\  
  \lbrack b_k\, ,\, b_{-k}\rbrack =-i ~,~ \lbrack b_k\, ,\, 
  b^{\dagger }_{k}\rbrack =0     &
\mbox{for $k^2+\mu^2<V_0$.}   
\end{array}
\nonumber   
\label{tcomm}
\eeqa
By using the linear transformation properties of hypergeometric functions
\cite{Abgun}, we have the following Bogolubov transformations between 
in and out modes.
\beq
u^{\rm in}_k(t,x)=\alpha_ku^{\rm out}_k(t,x) +\beta_ku^{\rm out\ast }_
{-k}(t,x)
\label{Btrans}
\eeq
where
\beq
\alpha_k \sim \frac{\Gamma (1-i\omega_{\rm in}/\rho )\Gamma 
(-i\omega_{\rm out}/\rho )}{\Gamma (-i\omega_+/\rho )\Gamma 
(1-i\omega_+/\rho )}, \qquad 
\beta_k \sim \frac{\Gamma (1-i\omega_{\rm in}/\rho )\Gamma 
(i\omega_{\rm out}/\rho )}{\Gamma i\omega_-/\rho )
\Gamma (1+i\omega_-/\rho )}. \nonumber 
\eeq
Hence, 
\beq
b_k = \alpha_k a_k +\beta^{\ast }_{-k} a^{\dagger }_{-k}. \nonumber 
\eeq
From Eq.~(\ref{tcomm}), we obtain
\beqa
\begin{array}{ll}
  |\alpha_k|^2-|\beta_k|^2 =1,\qquad  \alpha_k\beta^{\ast }_k-\alpha_{-k}
\beta^{\ast }_{-k}=0  & \mbox{for $k^2+\mu^2 > V_0$,}  \\
  |\alpha_k|^2-|\beta_k|^2 =0,\qquad  \alpha_k\beta^{\ast }_k-\alpha_{-k}
  \beta^{\ast }_{-k}=-i \qquad  & \mbox{for $k^2+\mu^2 < V_0$.}
\end{array}
\nonumber 
\eeqa
Notice that the second `converted' relations result from the unusual form 
of the commutation relations among mode operators for imaginary frequencies.

Since the scalar field becomes free as $t\to -\infty $, there exists 
a well-defined Fock representation near $t\sim -\infty $ as usual. Let
$|0\! >_{\!\! \rm in}$ be a vacuum state defined as 
$a_k|0\! >_{\!\! \rm in}=0$ for all $k$. As seen before, however, there is 
no well-defined vacuum state in the remote future due to the presence 
of imaginary frequency modes. Now let us consider a ``particle" detector 
linearly coupled to the field near $t\sim \infty $ placed in the in-vacuum 
state $|0\! >_{\!\! \rm in}$. The transition probability of the detector 
from $|E_0\! >$ to $|E'\! >$ with $E=E'-E_0$, in first order perturbation 
theory, is proportional to the response function ${\cal F}(E)$\cite{BD}.
\beq
{\cal F}(E)=\lim_{T\to \infty }\int^{T+\Delta T}_{T}dt 
\int^{T+\Delta T}_{T}dt'\,  
e^{-iE(t-t')} {_{\rm in}\!\! }<\! 0|\phi [x(t)]\phi [x(t')]|0\! >
_{\!\! \rm in}.     \nonumber   
\eeq
Since ${_{\rm in}\!\! }<\! 0|\phi [x(t)]\phi [x(t')]|0\! >_{\!\! \rm in}
=\sum_k u^{\rm in}_k(x)u^{\rm in\ast }_k(x') $ by using Eq.~(\ref{tdecom}),
we see
\beq
{\cal F}(E)=\lim_{T\to \infty }\sum_k\, |\int^{T+\Delta T}_{T}\! dt\, 
e^{-iEt}u^{\rm in}_k(x)\, |^2 .    
\eeq
From the Bogolubov transformations Eq.~(\ref{Btrans}) and the asymptotic 
behavior of $u^{\rm out}_k(x)$ near $t\sim \infty $, we obtain finally
\beqa
{\cal F}(E)\!\!\! &=&\!\!\! \lim_{T\to \infty }\sum_{k^2+\mu^2>V_0}
\frac{1}{4\pi \omega }\{ |\alpha_k|^2(\frac{\sin E_+\Delta T}{
E_+})^2 +|\beta_k|^2(\frac{\sin E_-\Delta T}{E_-})^2
\nonumber  \\
& &+\frac{1}{2}{\rm Re}[\alpha_k\beta^{\ast }_k\frac{\sin E_+\Delta T\sin 
E_-\Delta T}{E_+E_-}e^{-2i\omega (T+\Delta T/2)}]\}  \nonumber   \\
& &+\sum_{k^2+\mu^2<V_0(k>0)}\frac{1}{4\pi |\omega |}\{ |\frac{\sin E_+
\Delta T}{E_+}|^2 [(|\alpha_k|^2+|\beta_{-k}|^2)e^{2|\omega |(T+\Delta T/2)}
\nonumber   \\
& &+(|\alpha_{-k}|^2+|\beta_k|^2)e^{-2|\omega |(T+\Delta T/2)}]   
+\frac{1}{2}{\rm Re}[\alpha_k\beta^{\ast }_k(\frac{\sin E_+
\Delta T}{E_+})^2 +\alpha_{-k}\beta^{\ast }_{-k}(\frac{\sin 
E_-\Delta T}{E_-})^2]\}    \nonumber 
\eeqa
where $\omega =\omega_{\rm out}(k)=\sqrt{k^2+\mu^2-V_0}$ and $E_{\pm }= 
E\pm \omega_{\rm out}$ for real frequency modes and 
$\omega_{\rm out}(k)=\pm i\sqrt{V_0-(k^2+\mu^2)}$ and $E_{\pm }= 
E\pm i|\omega_{\rm out}|$ for imaginary frequency modes with 
$\pm |k|$ momentum, respectively. 

Noticing that $(\sin \varepsilon \Delta T/{\varepsilon })^2/
{\Delta T} \sim \delta (\varepsilon )$ for $\Delta T \gg 1$, 
the transition rate is given by
\beq
\frac{{\cal F}(E)}{\Delta T} \sim \sum_{k^2+\mu^2>V_0}
\frac{|\beta_k|^2}{4\pi 
\omega_{\rm out}}\delta (E-\omega_{\rm out}) +\sum_{k^2+\mu^2<V_0(k>0)}
\frac{|\alpha_k|^2+|\beta_{-k}|^2}{4\pi |\omega_{\rm out}|(E^2+|\omega |^2)}
\frac{e^{2|\omega |\Delta T}}{\Delta T}e^{2|\omega |T} \nonumber   
\eeq
in the limit of $T,\Delta T \to \infty $. This result shows non-vanishing 
excitations of the particle detector related to the imaginary frequency
modes as well as the usual contributions due to the positive energy mode
mixing in real frequency modes. Furthermore, the contributions related to
imaginary frequency modes grow exponentially in $T$, implying some 
instability in the future probably due to no boundedness below in the 
energy spectrum. Note also that the $\delta $-function in the 
first term implies the energy conservation, that is, at first order 
perturbation theory, only the real frequency field mode whose quantum energy 
is same as that of the particle detector ($E=\omega_{\rm out}$) can excite 
the particle detector. For imaginary frequency modes, however, all modes 
contribute to the excitation possibly because the energy spectrum for any
imaginary frequency mode is continuous as shown in Sec. \ref{QPIFM}.  

\section{Discussion}
\label{disc}

In this paper, we have shown why one needs to consider complex frequency
modes in investigating the vacuum stability for a scalar field near stars
with ergoregions, and how the quantization can be formulated in the
presence of imaginary frequency modes for a real scalar field interacting
with external potentials. As simple systems, we examined a one-dimensional
square-well potential and a scalar field with varying mass. We found that
it is possible to quantize the field in the presence of imaginary frequency
modes, but the mode operators for them do not admit a Fock-like 
representation  and so no particle interpretation. The Hamiltonian operator 
for imaginary frequencies are equivalent to a system of a set of 
two decoupled {\it repulsive} harmonic oscillators. Therefore, the energy 
spectrum is continuous, no normalizable energy eigenstate exists, 
and there is no ground state. The excitation of the particle detector 
placed in the in-vacuum state for the field with time varying mass has
contributions related to imaginary frequency modes as well. 
The transition rate, however, is not stationary but is exponentially 
increasing in time. 

For the issue of vacuum stability near rapidly rotating stars with 
ergoregions, these 
results strongly indicate that there exists the quantum instability 
as well corresponding to the classical ergoregion instability shown in 
the appendix. However, its character will be very different from the 
conventional analysis of the quantum vacuum instability in the case of 
a rotating black hole because the analysis in that case relies on the 
existence of vacuum states natural to two asymptotic regions 
in the past and the future or the existence of two equivalent complete 
bases of the Hilbert space whereas the inclusion of imaginary frequency 
modes does not admit a Fock-like representation or a vacuum state. 
Non-vanishing $\beta $-coefficients in the Bogolubov transformations 
do not give a direct
interpretation of particle creation for imaginary frequency mode 
operators, not only because the particle interpretation breaks down, but
also because any energy eigenstate for imaginary frequency modes is 
non-normalizable. For some cases, however, one can use Unruh's ``particle" 
detector model to extract some useful physics as we have shown in the
previous section for a scalar field with time varying mass. For general
cases, it would perhaps be preferable to take the point of view that 
the fundamental object in quantum field theory is the field operator
itself, not the ``particles" defined in a preferred Fock space\cite{FvsP,Wald}.
For example, the expectation value of the energy-momentum tensor should 
be a meaningful quantity. However, renormalization of the energy-momentum 
in the presence of complex frequency modes would have to be understood 
first. As far as we know, this interesting issue has never been addressed.  
A direct application of the present work to the issue of quantum instability
near a star with eroregion will appear elsewhere.  

It is also an interesting issue how to construct a quantum field theory
in the presence of complex frequency modes by using the algebraic approach.
To obtain a quantum description of fields in this approach, one first defines
the $\ast$-algebra of field operators and then constructs the Hilbert space 
of states by choosing an appropriate $\ast$-representation, equivalently a
suitable complex structure, of this $\ast$-algebra with a set of rules for
dynamics. The most difficult part in this prescription is to single out 
the `correct' representation among all possible $\ast$-representations. 
In Ref.~\cite{AM}, 
Ashtekar and Magnon have shown how certain physically motivated
requirements select a unique complex structure and so the `correct' 
representation. In the presence of complex frequency mode solutions,
however, it is not clear at present stage whether or not this algebraic 
approach is extendable. It is not only because the field grows 
unboundedly in time due to the imaginary part of the complex frequency
so that the very construction of the $\ast$-algebra itself would break down,
but also because the inner product becomes indefinite as shown in 
Eq.~(\ref{Qortho2}) so that it is unclear whether one can find a complex
structure compatible with the symplectic structure. 
Of course, the field would not blow up in the realistic
system. In other words, if we also include the dynamics of the external
system which is producing the external potential, then the field will stop
the exponential increase in time when the external potential loses its 
energy into the field and becomes smaller than a certain critical value,
for example, when $V_0<\mu^2$ in the case of square-well potential and
when the ergosphere disappears in the case of a rapidly rotating star as
mentioned by the authors in Ref.~\cite{AM}. 

There are many other fields in physics in which complex frequency modes
play important roles. Generically, if a system stores some ``free"   
energy which can be released through interactions, then some amplifications
occur, revealing complex frequency modes classically. In a system of plasma,
for instance, a small perturbation of electric field exponentially 
increases in time if the phase velocity of the perturbed field is smaller
than the velocity of charged particles, and is damped in the opposite
case. The energy stored in plasma is released quickly by a small 
perturbation, giving complex frequency modes\cite{plasma}. In a tunable 
laser, the energy stored in dielectric material amplifies an incident
light and results in the intensity increasing of the output laser beam.
In a field theoretic treatment of the system, the dielectric material
plays the role of a source producing an external potential and 
it is possible for complex 
frequency modes to occur under suitable conditions\cite{Raj}. 
In the theory of linear quantum amplifiers\cite{lamp}, one assumes a time 
dependent annihilation operator, $a(t)=a(0)e^{Wt/2-i\omega t}$ with a gain 
factor $W$. This gain factor $W$ may be interpreted simply as coming from 
the imaginary part of a complex frequency mode in the second quantization 
scheme where one does not need to assume the non-unitary evolution of the 
mode operator. In addition, the repulsive harmonic oscillator representation 
appeared in the quantization formalism in this paper may be able to explain 
the model of the inverted oscillator amplifier\cite{inoscamp} and the 
initial stages of the superfluorescence process modeled by Glauber and 
Haake\cite{GH} in a more fundamental sense.     
Therefore, the quantization formalism described at the present work may be 
useful to understand those phenomena in the context of quantum field theory.

\vskip 1cm
The author would like to acknowledge useful discussions with A. Ashtekar,  
S. Corley, B.R. Iyer, C.W. Misner, P. Mohanty, R. Nityananda, J. Samuel, 
and J.H. Yee. Especially, I am deeply indebted to Ted Jacobson for many 
helpful discussions and suggestions. This research was supported in part 
by NSF Grant PHY94-13253.

\vskip 1cm
\centerline{\Large {\bf Appendix: Occurrence of Complex Frequency Modes}}
\vskip 0.5cm

In this appendix we will basically follow Vilenkin's demonstration showing
the generation of exponentially amplified waves, which is related to 
the occurrence of complex frequency modes, in the background metric of rapidly 
rotating stars. The physical interpretation of the demonstration will 
be the following. An incoming radial wave {\it in a superradiant mode} will 
be scattered by the potential barrier near the ergosurface with a reflection 
coefficient {\it bigger} than unity. The transmitted wave will be trapped 
inside the ergoregion by passing through the center of the rotating star 
and by being reflected at the ergosurface repeatedly. At each reflection 
at the ergosurface, the transmitted waves will be amplified and escape to 
infinity. Thus, the total energy of outgoing waves looks like an exponential 
amplification of that of the initial ingoing wave, reminiscent of laser action. 

The system we consider is a massless real scalar field $\phi (x)$ 
satisfying the Klein-Gordon equation
\beq
\Box \phi = \frac{1}{\sqrt{-g}}\partial_{\mu }(\sqrt{-g}g^{\mu \nu } 
\partial_{\nu }\phi ) = 0
\label{KG1}
\eeq
in the spacetime of a rotating star {\it with ergosphere but no event 
horizon}. Thus, the system is stationary and the spacetime outside the
surface of the rotating star is described by the Kerr metric. The inside 
of the star body is assumed to be regular. We assume that the surface of 
the star body is near the outside of the corresponding event horizon,  
if it would have existed, of the Kerr metric. 

Since the Kerr metric has two Killing vector fields(i.e., $\xi^a =
(\partial /{\partial t})^a $ and $\psi^a =(\partial /{\partial \varphi }
)^a $ in Boyer-Linquist coordinates $(t,r,\theta ,\varphi )$) 
and Carter's constant from $\nabla_{(a}K_{bc)}=0$, Eq.~(\ref{KG1}) is 
separable and admits a complete set of solutions of the form\cite{Unford,MDO} 
\beq
\phi (x) = e^{-i\omega t + im\varphi } U_{\omega lm}(r) S_{\omega lm}
(\theta ).
\label{sep}
\eeq
Then, Eq.~(\ref{KG1}) becomes, outside the rotating star body,  
\beq
\frac{\partial }{\partial r}(\triangle \frac{\partial \phi }{\partial r}
)+ \frac{1}{\sin \theta }\frac{\partial }{\partial \theta }(\sin  
\theta \frac{\partial \phi }{\partial \theta }) - m^2(\frac{1}{\sin^2
\theta }-\frac{a^2}{\triangle })\phi  \nonumber   \\
 -4am\omega \frac{Mr}{\triangle }\phi +\omega^2 [\frac{(r^2+a^2)^2}{
\triangle }-a^2\sin^2\theta ]\phi =0 
\label{KG2}
\eeq
where $\triangle = r^2-2Mr+a^2 $, and $M$ and $a$ are the total mass of 
the rotating star and the angular momentum per unit mass, respectively. 
$l$ and $m$ are integers satisfying $|m| \leq l $. Defining 
$U_{\omega lm}(r) = R_{\omega lm}(r)/{(r^2+a^2)^{1/2}} $, the radial part 
of Eq.~(\ref{KG2}) yields
\beq
\frac{d^2R_{\omega lm}}{dr^{\ast 2}} - V_{\omega lm}(r)R_{\omega lm}
=0 
\label{rKG}
\eeq
where the $r^{\ast }$ is a ``generalized" tortoise coordinate defined
by $dr^{\ast }/{dr} =(r^2+a^2)/{\triangle } $. Since $\triangle (r) 
\rightarrow 0 $ as $r$ approaches the horizon radius $r_{+} =M+\sqrt{
M^2-a^2} $, one can easily see $r^{\ast } \rightarrow -\infty $ as 
$r$ decreases to $r_{+}$. 

The asymptotic behavior of the effective potential $V_{\omega lm}(r)$ 
induced through the interaction with the gravitational field is as follows:
\beq
V_{\omega lm}(r) \sim \left\{\begin{array}{ll}
                               -\omega^2  & 
\mbox{as $r^{\ast }\rightarrow \infty $}    \\ 
                               -(\omega -m\Omega_H)^2 \qquad   &
\mbox{as $r^{\ast }\rightarrow r_0^{\ast }$.} 
			     \end{array}
                      \right.
\label{pot}
\eeq
Here the $r_{0}^{\ast }$ corresponds to $r=r_0$ ( $\stackrel{>}{\sim }r_+ $ 
and so $r_{0}^{\ast } \sim  -\infty ) $ at which the surface of the 
rotating star is located. $\Omega_H$ is defined as 
$\Omega_H =a/(2Mr_+)$, 
which is the angular velocity of the horizon of the Kerr metric.
Note that, between two asymptotic regions, there exists 
a potential hump which grows as $l$ increases. Note also
that the left asymptotic value $-(\omega -m\Omega_H)^2$ varies from 
$0$ to $-\infty $ as $m\Omega_H $ changes. In particular, a deep 
potential well is produced for a big value of $m\Omega_H$ which 
leads to the classical superradiance mode. 

The behavior of this potential will not change much when it crosses 
the surface of the rotating star body. However, we put a totally
reflecting mirror on the surface of the star. This reflection 
boundary condition is justified if the scalar field does not interact 
with the matter of the rotating star body. That is, an ingoing 
spherical wave will cross the surface of the star, pass the center, 
and go back to the outside of the star without being changed much.
We achieve this reflection boundary condition on the star surface
by setting an infinite potential wall at $r^{\ast }=r_{0}^{\ast }$.
Therefore, any scalar field satisfying this reflection boundary 
condition will vanish at the reflecting mirror, i.e., $\phi (x) =0$ at
$r^{\ast }=r^{\ast }_0$. In addition to this inside boundary condition, 
we also assume that the field does not blow up at infinity.

Now, the question arises as to whether or not there exist complex
frequency modes in the geometry described above. In the case of a 
rotating black hole, Detweiler and Ipser\cite{Dipser} have made a 
numerical search showing that complex frequency modes of Im $\omega > 0$ 
do not exist. After many years, Whiting\cite{Whiting} finally proved 
this analytically for massless fields. 
On the other hand, in the case of an ultrarelativistic rotating star 
where an ergosphere appears and a reflecting mirror boundary surface is 
assumed, the occurrence of complex frequency modes has been shown by 
Vilenkin\cite{vilen}. He basically shows how the existence of the 
reflection boundary condition produces the exponential amplification of 
the waves in superradiant modes. The issue on the ergoregion instability 
has also been studied by many authors from various points of 
view\cite{ergoinst}. 

Let us first consider the rotating black hole case. From the asymptotic 
form of the potential in Eq.~(\ref{pot}), we can easily construct the 
linearly independent solutions $R^{\pm }_{\omega lm}$ of Eq.~(\ref{rKG}) 
whose asymptotic forms are 
\beq
R^{+}_{\omega lm} \sim \left\{\begin{array}{ll}
                                B^+e^{-i\tilde{\omega }r^{\ast }}   & 
\mbox{as $r^{\ast } \rightarrow -\infty  $}    \\ 
         e^{-i\omega r^{\ast }} + A^+e^{i\omega r^{\ast }} \qquad   &
\mbox{as $r^{\ast }\rightarrow \infty $}   
                              \end{array}
                       \right.
\label{+sol}
\eeq
and
\beq
R^{-}_{\omega lm} \sim \left\{\begin{array}{ll}
                                e^{i\tilde{\omega }r^{\ast }} + A^-e^{
                                -i\tilde{\omega }r^{\ast }} \qquad   &
\mbox{as $r^{\ast } \rightarrow -\infty $}     \\   
                                B^-e^{i\omega r^{\ast }}    &
\mbox{as $ r^{\ast } \rightarrow \infty $} 
			      \end{array}
			\right.
\label{-sol}
\eeq
where $\tilde{\omega }=\omega -m\Omega_H $ and $\omega >0 $. As one
can see, $R^{\pm }_{\omega lm} $ are the waves originating at infinity 
and at the horizon, respectively. From Eq.~(\ref{rKG}), we obtain
\beq
\frac{d}{dr^{\ast }}(R_1\frac{dR_2}{dr^{\ast }} - R_2\frac{dR_1}
{dr^{\ast }}) = (V_2-V_1)R_1R_2 
\label{Wron1}
\eeq
for any two solutions $R_{\omega_1l_1m_1} $ and $R_{\omega_2l_2m_2}$. 
When $V_2=V_1$, the Wronskian ($=R_1R^{\prime }_2 -R_2R^{\prime }_1$)
is constant, and then it leads to the relations
\beqa
1-|A^+|^2 = \frac{\tilde{\omega }}{\omega }|B^+|^2 , \qquad 
1-|A^-|^2 = \frac{\omega }{\tilde{\omega }}|B^-|^2 ,    \nonumber  \\
\omega B^- = \tilde{\omega }B^+, \qquad A^{+\ast }B^- = -\frac{\tilde{
\omega }}{\omega }A^-B^{+\ast }.
\label{Wron2}
\eeqa
It can be easily seen from the above equations that $|A^+|^2 > 1$ and
$|A^-|^2 > 1$ for $\tilde{\omega } < 0$(i.e., $\omega < m\Omega_H$), 
which are the so-called superradiant modes. This property indicates that, 
if an ingoing wave packet sharply peaked in frequency in the range of
superradiance is scattered off the black hole, an amplified fraction 
$|A^+|^2 $  will be reflected back to infinity with being amplified and a 
fraction $|A^+|^2 -1 =-\frac{\tilde{\omega }}{\omega }|B^+|^2 $ will be 
transmitted through the ergoregion and absorbed into the event horizon
finally. 

In the case of a rotating star, however, this transmitted wave becomes an 
outgoing wave after passing through the center of the star and will then 
scatter at the potential barrier in the ergoregion. The reflected fraction 
this time will be $|A^-|^2 \, (>1)$ and 
$|A^-|^2 - 1= -\frac{\omega }{\tilde{\omega }}|B^-|^2$ for the transmission. 
This process will be repeated presumably until all rotational energy 
of the star is released to infinity. Assuming the energy of the incident 
wave to be $E_0$, the total energy escaped to infinity is 
\beq
E_{\rm OUT} = E_0(|A^+|^2 +|B^-B^+|^2\sum^{\infty }_{n=0} 
|A^-|^{2n} ).
\label{Eout}
\eeq
The energy accumulated in the ergoregion equals
\beq
E_{\rm IN} = E_0 \frac{\tilde{\omega }}{\omega }|B^+|^2\lim_{ 
n\to \infty } |A^-|^{2n} .
\label{Ein}
\eeq
For non-superradiant modes $\tilde{\omega } >0$, the above two equations
are still satisfied and we find $E_{\rm IN} =0$ and $E_{\rm OUT}=E_0$ since 
$|A^-| < 1$ and $\sum^{\infty }_{n=0} |A^-|^{2n} =(1-|A^-|^2)^{-1} $. 
On the other hand, if $\tilde{\omega }<0$, i.e., superradiant modes, then 
$|A^-|>1$ and we find both $E_{\rm IN}$ and $E_{\rm OUT}$ diverge. 
However, the sum is always the incident energy, 
$E_{\rm IN} + E_{\rm OUT} = E_0 $. 
If $t_0$ is the time interval between two 
successive reflections at the mirror $r^{\ast }=r^{\ast }_0 $, then the 
radiated power measured at infinity is 
\beq
\frac{dE}{dt} \sim |A^-|^{2t/{t_0}} = e^{t/{\tau }} 
\label{power}
\eeq
where the ``e-folding" time is $\tau =t_0/{2\ln |A^-|}$. 
Assuming $t_0\sim 2|r^{\ast }_0|/c$,
\beq
\tau \simeq |r^{\ast }_0|/{\ln |A^-|}.
\label{efold}
\eeq
Note that $\tau \to \infty $ as $\tilde{\omega } \to 0$, i.e., $|A^-|\to 1$, 
and so the amplification of waves becomes very weak.

This effect of exponential amplification in the presence of the reflection
boundary strongly suggests that there exist unstable mode solutions of 
Eq.~(\ref{KG1}) which grow exponentially in time. In fact, there exists 
such an unstable mode solution in the presence of the reflection boundary 
condition which is purely outgoing at infinity like Eq.~(\ref{-sol}) up to 
an overall scaling. Since such a mode solution should not blow up at 
infinity(in order to be included in standard approach to quantization), 
it must satisfy the following two boundary conditions, 
including the one on the surface of the star,
\beq
R_{\omega lm}(r^{\ast }=r^{\ast }_0)=0 ; \qquad |R_{\omega lm}(
r^{\ast }\to \infty )| < \infty .
\label{bc}
\eeq
From the first condition, we find
\beq
e^{i\tilde{\omega }r^{\ast }_0} + A^-e^{-i\tilde{\omega }r^{\ast }_0}
=0.
\label{INbc}
\eeq
Thus, defining $A^- =|A^-|e^{i\delta } $,
\beq
\omega = m\Omega_H +\frac{(2n-1)\pi -\delta }{2r^{\ast }_0} +i \frac{
\ln |A^-|}{-2r^{\ast }_0}.
\label{Cfreq}
\eeq
We find that the imaginary part of the frequency $\omega $ is {\it non-zero:} 
\beq
\omega_I =\hbox{Im} ~\omega =\ln |A^-|/{(-2r^{\ast }_0)}.
\label{Im}
\eeq
If $|A^-| <1$, then $\omega_I <0$ and we see the boundary condition at 
infinity in Eq.~(\ref{bc}) is not satisfied. Thus, only the ``superradiant" 
modes ($|A^-|>1$) as viewed from inside the ergosurface give satisfactory 
solutions. Since the time dependence of the unstable mode solutions are 
$\phi_{\omega } \sim e^{-i\omega t}=e^{-i\omega_Rt}e^{\omega_It}$, the 
radiated power to infinity will be proportional to 
$e^{2\omega_It}=e^{t/{\tau }}$, and thus we obtain 
$\tau =(2\omega_I)^{-1}=-r^{\ast }_0/{\ln |A^-|}$, which agrees with 
Eq.~(\ref{efold}). 

According to Eq.~(\ref{Im}) $\omega_I$ diverges as $|A^-|$ ($>1$) increases. 
It, however, can be shown that $\omega_I$ is bounded. By multiplying 
Eq.~(\ref{KG2}) by $\phi^{\ast }$ and integrating over $r$ and $\theta $, 
we obtain 
\beqa
\int^{\infty }_{r_0}\!\! \int^{\pi }_{0}\! \sin \theta drd\theta \!
&\{ &\!\! \triangle |\frac{\partial \phi }{\partial r}|^2+|
\frac{\partial \phi }
{\partial \theta }|^2+[\frac{m^2}{r^2\sin^2\theta }(\triangle -a^2 
\sin^2\theta )+\frac{4amM}{r}\omega      \nonumber   \\
& &-\frac{\omega^2}{r^2}((r^2+a^2)^2-a^2\triangle \sin^2\theta )]
\frac{r^2}{\triangle }|\phi |^2 \} =0     \nonumber 
\eeqa
since the boundary terms vanish at $r=r_0\, ,\, \infty $ and $\theta 
=0\, ,\, \pi $. First notice that $\omega $ must be real for $m=0$ in the 
above equation. Using $[(r^2+a^2)^2-a^2\triangle \sin^2\theta ]/{r^2} 
\geq r^2_+ +a^2+2Ma^2/{r_+}$ for $r\geq r_+$, we see
\beq
m\omega_R \geq 0, \qquad \qquad 0\leq |\omega_R|\leq |m|\Omega_H  
\nonumber    \\
\eeq
and 
\beq
\omega^2_I \leq (1+a^4/{r^4_+})\omega^2_R+2mM(m+2a\omega_R)/{r^3_+}.
\nonumber
\eeq
Therefore, unstable eigenfrequencies are confined in a bounded region.

We have seen that unstable modes with complex frequencies occur in the 
region of superradiance for a massless scalar field when a reflection
boundary condition is assumed somewhere inside the ergoregion, which 
can be regarded as a rapidly rotating star. In the case of rotating 
black holes, no such complex frequency modes exist.
For a small negative energy perturbation in the ergoregion
is simply absorbed by the event horizon so that the outgoing positive 
energy flow is not amplified any more. In contrast, for a rotating
star, the negative energy is trapped inside the ergoregion giving 
amplified positive energy flows to infinity repeatedly, and so 
revealing an {\it exponential} amplification. Of course, the negative 
energy inside amplifies itself to conserve energy. This process leads to 
instability and gives unstable modes whose frequencies are complex.



\begin{thebibliography}{99}
\bibitem
{note} This assumption has indeed been proved by Whiting\cite{Whiting}.
\bibitem
{lcavity} In fact, the one-dimensional Maxwell wave equation $[n^2(z)
          \partial_t^2-\partial_z^2]E(z,t)=0$\cite{LSL} can be related to 
          Eq.~(\ref{KGM}) for massless fields in one dimension by a
          transformation $E(z,t)=n^{-1/2}(z)\phi (x,t)$ and $dx/{dz}=n(z)$ 
          with the potential $V(x)=-n^{-3/2}\partial_z^2n^{-1/2}$\cite{SO}.
\bibitem
{footnote} In a Hydrogen atom, the energy spectrum from Dirac's equation
indeed shows that the ground state energy becomes imaginary as the atomic
number $Z$ is larger than $\alpha^{-1}\simeq 137$, e.g., $E_{\rm g}=mc^2
\sqrt{1-(Z\alpha )^2}$.
\bibitem
{BC} By suitable boundary behavior, we mean that the spatial part of 
     solutions has a compact support, is periodic, or goes to zero 
     sufficiently fast as one approaches spatial infinity.
\bibitem
{foot2} Since we are assuming a non-singular potential well around 
	$x=\pm a$, the first derivative of $\phi_j(x)$ must be continuous.
\bibitem
{Raj} Private discussion with Pritiraj Mohanty at the university of 
      Maryland. He has been working on a system of cavity filled with 
      dielectric material and found also that the study of the role of 
      complex frequency modes in quantum field theoretic regime is 
      crucial  to understanding the related optical phenomena.
\end{thebibliography}
\end{document}